\documentstyle[a4,12pt,psfig]{article}
\def\epsfig{\psfig}
\newcommand{\Lie}[1]{{\cal L}_{#1}}
\begin{document}
\begin{titlepage}
\begin{center}
\hspace*{10cm} TUM-HEP-257/96\\
\hspace*{10cm} LMU-TPW 96-26\\
\hspace{10cm} hep-th/9610062\\
\hspace{10cm} October 1996\\
\vspace*{1cm}

{\LARGE Non Abelian T-duality for open 
strings\footnote{Talk given by S.\ F\"orste at the International Symposium
on the Theory of Elementary Particles Buckow, August 27-31, 1996;\\
Work partly supported
by the EC programme SC1-CT92-0789 and the European Commision TMR
programmes ERBFMRX-CT96-0045 and ERBFMRX-CT96-0090}}\\

\vspace{1.2cm}

{\large Stefan F\"orste$^{\S}$
\footnote{ E-Mail:Stefan.Foerste@physik.uni-muenchen.de}, 
Alexandros A.\ Kehagias$^{\sharp}$
\footnote{E-Mail: kehagias@physik.tu-muenchen.de}
 and
Stefan Schwager$^{\S}$
 \footnote{E-Mail:Stefan.Schwager@physik.uni-muenchen.de}}\\
\vspace{.8cm}

${}^{\S}$
{ Sektion Physik\\
Universit\"at M\"unchen\\
Theresienstra\ss e 37, 80333 M\"unchen\\
Germany}\\
\vspace{.6cm}

${}^\sharp${ Physik Department \\
Technische Universit\"at M\"unchen\\
D-85748 Garching, Germany}\\
\end{center}
\vspace{1cm}

\begin{center}
{\large Abstract}
\end{center}
\vspace{.3cm}
In the first part of the talk we discuss T-duality for a free boson on a 
world sheet with boundary in a setting suitable for the generalization to
non-trivial backgrounds. The gauging method as well as the canonical 
transformation are considered. 
In both cases Dirichlet strings as T-duals of Neumann strings arise in
a generic way.
In the second part the gauging method is employed to construct the T-dual
of a model with non-Abelian isometries.
\end{titlepage}
\newpage
\section{Introduction}
During the last years the appealing picture arose that all string
theories are just suitable descriptions of one underlying
unique theory. This picture is supported by the discovery
that many string theories are linked to each other via duality
transformations. Among those dualities T-duality is around already 
for quite some time \cite{kikawa,buscher}. It is a weak coupling/weak coupling
duality relating small compactification radii on one side to
large compactification radii on the dual side. In some cases
it maps different looking string theories on each other like
type IIA on type IIB \cite{polch,dine,berg,alex} or the heterotic $E_8\times E_8$
on a heterotic $SO(32)$ string (with symmetry breaking 
Wilson lines)\cite{ginsparg}.  
Already  some time ago it has been observed that Dirichlet boundary 
conditions arise in the
T-dual description of toroidally compactified
open strings with the usual Neumann boundary 
conditions \cite{polch,horava,leigh}. 
That is the T-dual picture includes extended objects (D-branes) in
a natural way,  (for an
excellent review see \cite{notes}).
Those D-branes play an important role in establishing string dualities
like e.g.\ type I/heterotic duality \cite{jo-witten} or  in identifying
heterotic/heterotic duality in six dimensions with T-duality
in an open string theory \cite{berkooz}.
Therefore it appears to be interesting to generalize the T-duality
transformation 
for open strings living in a non trivial background.

That in  backgrounds possessing 
a Poisson-Lie symmetry T-duality maps Neumann on Dirichlet
conditions has been shown in \cite{klimcik}.  Later on,  T-duality was 
carried out in general backgrounds with Abelian
isometries \cite{alv,do}.  
The generalization for backgrounds with
non-Abelian 
isometries is worked out in \cite{us}, which will be partly reported
in the present talk. 
There have been some publications discussing T-duality 
corresponding to non-Abelian isometries 
\cite{quev}, \cite{alot1,alot2,alot3,alot4,alot5,grp}. Here we will follow the 
initial work of
\cite{quev}, i.e.\ we have to restrict ourself on semi-simple isometry
groups. 

Non-Abelian T-duality for open strings has been discussed in \cite{yolanda}
as a canonical transformation. In \cite{yolanda} results differing from ours
have been obtained. Below we will argue that a proper inclusion of 
boundary effects in the canonical transformation gives Dirichlet
conditions as a dual for free varying ends of the string, generically.

In the next section we will discuss T-duality for the simplest example,
{\it viz}.\ a free boson. We will use the gauging method \cite{ro-ve} and we 
will also discuss a canonical transformation approach. Knowing the results
for the closed string it will be straightforward to extend those considerations
to less trivial backgrounds. In section three we gauge the non-Abelian
isometries of a sigma model with boundary and sketch the T-duality
transformation. The last section concludes the talk also mentioning some
open problems. In a brief appendix we give basic ideas about the canonical
transformation.
\section{The free boson}
Combining the following discussion of the free boson with known results of
closed string theory it will be straightforward to modify closed string results
to the open string. The action for the free boson is given by
\begin{equation}
S = \int_{\Sigma} d^2z\, \partial_a X \partial^a X.
\end{equation}
In addition we specify boundary conditions
\begin{equation}
b^a\partial_a X_{| \partial \Sigma} = 0,
\end{equation}
with $b$ being some two dimensional vector. If $b$ is tangent to the boundary
we have, up to a constant, Dirichlet conditions and otherwise Neumann or
mixed conditions. This model is invariant under global shifts of $X$.
Now we are going to gauge that symmetry and at the same time to suppress
excitations of the gauge fields, i.e. to rewrite the theory to an equivalent one.
Integrating out the gauge fields will then give the T-dual model.
When suppressing gauge field excitations one also has to care about global
excitations in order to obtain a full equivalence. That issue will be neglected 
here since in the non-Abelian case this is an unresolved problem also
in the closed string \cite{alot1,alot2,alot3}. We want to construct an action 
invariant under
\begin{equation}
X \rightarrow X + f(z) 
\end{equation}
in combination with
\begin{equation}
\Omega_a \rightarrow \Omega_a - \partial_a f .
\end{equation}
Defining covariant derivatives via
\begin{equation}
D_a X = \partial_a X + \Omega_a .
\end{equation}
a gauge invariant action is given by
\begin{equation}
S_{gauged} = \int_{\Sigma} d^2z\, \left( D_a X D^a X + \lambda F\right)
+\oint_{\partial \Sigma} 
ds\, \left( c t^a +\kappa (s) b^a\right)\Omega_a ,
\end{equation}
where $F$ is the field strength corresponding to the isometry gauge 
field $\Omega$, 
\begin{equation}
F = \epsilon^{ab}\partial_a \Omega_b,
\end{equation}  $\lambda$ is a Lagrange 
multiplier forcing the field strength $F$ to vanish, $t$ is the tangent vector
on the 
boundary $\partial \Sigma$ and $c$ is an arbitrary constant. The second 
Lagrange multiplier $\kappa$ forces the $b$-component  of the gauge
field to vanish at the boundary. Integrating out $\lambda$ will constrain
the gauge fields to be pure gauge, integrating out $\kappa$ will imply 
boundary conditions on the gauge parameter such that the gauge
parameter can be absorbed by a shift in $X$ and the equivalence to the
ungauged model is established. 

\begin{figure*}
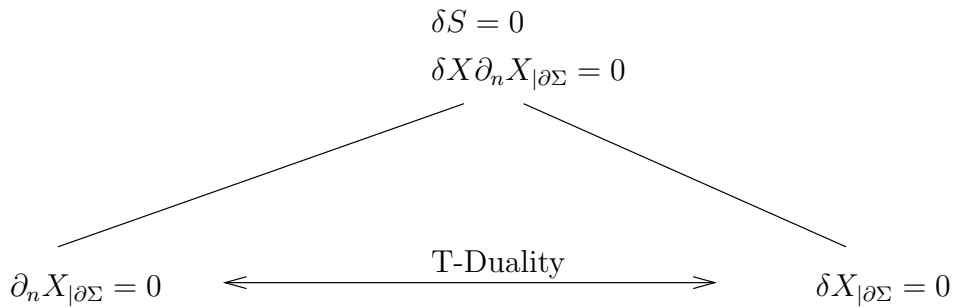
  
\begin{center}
\input buckow.pstex_t
\end{center}
\caption{T-duality for open strings interchanges free varying ends with 
fixed ends.}
\label{bildchen}
\end{figure*}

Before integrating out the gauge fields it is useful to employ partial 
integrations such that the action does not contain derivatives of $\Omega$. 
Then the $\Omega$ integral is ultra local and factorizes into an
integral where the argument of the gauge field is in the bulk times an integral
where the argument of the gauge field is on the boundary. The first one gives
the dual action in terms of the dual coordinate $\lambda$ whereas the later
integration results in a two dimensional delta function
\begin{equation} \label{delta}
\delta^{(2)} \left( ct_a + \kappa b_a + \lambda t_a\right).
\end{equation}
In the case that $b$ is not tangent to the boundary (i.e.\ Neumann or
mixed conditions) the second Lagrange multiplier $\kappa$ has to vanish
and the remaining one dimensional delta function imposes Dirichlet 
conditions on the dual coordinate
\begin{equation}  \label{dirich}
\lambda_{|\partial\Sigma} = - c.
\end{equation}
On the other hand if we started with Dirichlet conditions (up to a constant),
i.e.\ $b$ is tangent to the boundary, (\ref{delta}) just relates the first to the
second Lagrange multiplier on the boundary and no boundary conditions
for the dual coordinate are specified, 
i.e.\ the ends of the dual string vary freely,
(the infinite factor $\delta (0)$ can be
thought of as arising from specifying Dirichlet conditions only up to a 
constant). 
Finally we end up with the picture drawn in figure \ref{bildchen}.

Before considering non trivial backgrounds we would like to discuss how
to obtain the very same result as a canonical transformation. Now, let us
choose as a specific world sheet a strip in the upper half plane 
bounded by the $\sigma = 0$ and the $\sigma =\pi$ axis. The time $\tau $ is the
coordinate parallel to the boundary. Furthermore, let us assume that the
time asymptotic of the fields is such that the boundary can be closed at 
infinite times. 
The action for a free boson $X$ is given by
\begin{equation}
S= \int d^2\sigma \frac{1}{2}\left[ \left(\partial_{\tau}X\right)^2 - \left( 
\partial_\sigma X\right)^2
\right] 
+ \frac{1}{2}\left(\int_{\sigma = 0} -\int_{\sigma =\pi}\right)d\tau c 
\partial_\tau X,
\end{equation}
with $c$ being some arbitrary constant.
The boundary term is a total derivative integrated over a closed path and
hence vanishes. It will prove useful to keep it, nevertheless.
The canonical momentum is defined by
\begin{equation}
P = \frac{\delta S}{\delta \partial_\tau X},
\end{equation} 
leading to
\begin{equation}  \label{p-bulk}
P = \partial_\tau X \,\,\, \mbox{for\ } \sigma \in (0,\pi),
\end{equation}
\begin{equation} \label{p-bound}
P = \left\{ \begin{array}{lll}
\frac{c}{2} &\mbox{for}& \sigma = 0 \\ 
-\frac{c}{2} &\mbox{for}&\sigma =\pi \end{array}\right .
\end{equation}
for free varying ends, i.e.\ the net momentum flow through
the boundary is zero.
The generating functional for the canonical transformation\footnote{Some 
basics about canonical transformations are given in the appendix.}  
is given by the expression
\begin{equation}
F\left[ X , \tilde{X} \right] = \frac{1}{2}\int_0^\pi d\sigma \left( \tilde{X}
\partial_\sigma  X - X\partial_\sigma \tilde{X}\right)
\end{equation}
from which one obtains the canonical transformation via
\begin{equation}
\delta F = P \delta X - \tilde{P}\delta\tilde{X}.
\end{equation}
with
\begin{equation} \label{vari}
\delta F = \int_0^\pi d\sigma \left( - \delta X \partial_\sigma \tilde{X} +
\delta\tilde{X}  \partial_\sigma X\right) 
+\frac{1}{2}\left( \delta X \tilde{X} 
-X\delta \tilde{X}\right)_{\sigma = 0}^{\sigma = \pi}
\end{equation}
and (\ref{p-bulk})
we get
\begin{equation}
\begin{array}{ll}
\partial_\sigma \tilde{X}& = -\partial_\tau X \\
\tilde{P}& = -  \partial_\sigma X \end{array}\,\,\,  \mbox{for\ } \sigma \in (0,
\pi).
\end{equation} 
Eq.\ ({\ref{p-bound}) together with (\ref{vari}) gives 
\begin{equation}
\tilde{X} = - c \,\,\,\mbox{for\ } \sigma = 0,\pi
\end{equation}
which in turn freezes the $\tilde{X}$ variation to vanish at the boundary
and leaves the boundary value of the dual momentum open, i.e.\ we see
already at this stage that a Dirichlet string arises in the dual picture,
which will turn out to be a generic result as long as the original string has
free varying ends. (Note however that adding additional
boundary contributions to the generating functional
will change that result.)

\section{The gauged model and non-Abelian T-duality}
In this section we are going to gauge a model with non trivial
background following \cite{hull,osborn} and to construct the
T-dual model in analogy to the previous discussion.
The sigma model under consideration is
given by
\begin{eqnarray}
S&=&-\frac{1}{2}
\int_\Sigma d^2\sigma G_{mn}\partial_a X^m\partial^aX^n \nonumber\\
&&-\int_{\partial\Sigma} ds A_m\partial_sX^m+ 
\frac{1}{4\pi}\int_\Sigma d^2\sigma\ \sqrt{\gamma}
R^{(2)}\Phi -\frac{1}{2\pi}\int_{\partial\Sigma} ds k \Phi \, .
\label{openaction}
\end{eqnarray}
The non trivial fields are the target space metric $G$ the dilaton $\Phi$
(coupling to the scalar curvature $R^{(2)}$ in the bulk and to the
geodesic curvature $k$ on the boundary), and the U(1) gauge field $A$
coupling to the boundary. (The antisymmetric tensor field is taken to be trivial
for simplicity.)
Further let us assume that there are isometries forming a semi-simple
Lie group,
\begin{equation}
[\xi_I,\xi_J]={f_{IJ}}^K\xi_K\, .
\end{equation}
Then the coordinates $X^m$ transform as
\begin{equation}
\delta X^m=\epsilon^I\xi^m_I. \label{dtra}
\end{equation}
The conditions that the action (\ref{openaction}) is invariant under these
(global) transformations is that the Lie derivative of the dilaton vanishes and
that the Lie derivative of the U(1) gauge field is a pure gauge
\begin{equation}
\Lie{\xi_I}A_m =\partial_m\phi_I \, . \label{Lie}
\end{equation}
Note that the right hand side of (\ref{Lie}) is invariant under constant
shifts of $\phi_I $. From now on we will neglect the 
dilaton and comment on it in the conclusions.
By evaluating 
$\Lie{[\xi_I,\xi_J]}A_m$ we find that
 $\phi_I$ must satisfy the
consistency conditions
\begin{equation}
\Lie{\xi_I}\phi_J- \Lie{\xi_J} \phi_I = {f_{IJ}}^K\phi_K-k_{IJ}\, ,\label{consis
tency}
\end{equation}
where $k_{IJ}$ are constants transforming under the above mentioned 
constant shifts of $\phi_I$. 
The gauged model is obtained by introducing isometry gauge fields $\Omega$
and  demanding  invariance of the
action under 
\begin{eqnarray}
\delta X^m&=&\epsilon^I\xi^m_I \, , \nonumber \\
\delta \Omega_a^I&=& \partial_a \epsilon^I+{f_{KJ}}^I \Omega^K_a \epsilon^J \, .
\label{gtrans}
\end{eqnarray}
Since we have no antisymmetric tensor field in the bulk the gauge invariant
bulk part of the action is obtained by replacing partial derivatives with
covariant ones
\begin{equation}
D_aX^m = \partial_a X^m -\xi^m_I\Omega_a^I \, . \label{der}
\end{equation}
To construct an action with an invariant boundary term we employ
Noether's method, i.e.\ we add 
\begin{equation}
S^{(1)}=\int ds\, C_I\, \Omega^I_s
\end{equation}
and determine the fields $C_I$ such that the full action is gauge
invariant. That leads to two equations (for details see \cite{us})
\begin{eqnarray}
C_I&=&-A_m\xi^m_I+\phi_I \label{c1}+\lambda_I\\
\Lie{\xi_I}C_J&=&-{f_{JI}}^KC_K\, , \label{c2}
\end{eqnarray}
with $\lambda_I $ being some constant. These equations are compatible
provided that
\begin{equation} \label{kkk}
k_{IJ} -{f_{IJ}}^{K}\lambda_K =0\, ,
\end{equation} 
implying an integrability condition on the constants $k_{IJ}$
\begin{equation}
{f_{IJ}}^Kk_{KL} + cycl.\, perm.\, = 0.
\end{equation}
The gauged model is finally given by
\begin{equation} \label{ginv}
S=-\frac{1}{2}\int_\Sigma d^2\sigma G_{mn}D_a X^mD^aX^n \\
-\int ds\, \left(A_m\partial_sX^m-C_I\Omega^I _s\right) \, . 
\end{equation}

Now, one can perform a T-duality transformation for the considered model
in the same way as described in the beginning for the free boson. The details
are given in \cite{us}. In addition to the coordinates $X^m$ there might be 
additional coordinates $X^\alpha$ transforming as singlets under the
isometry. So, we add a Lagrange multiplier constraining the isometry
gauge field strength to vanish. This Lagrange multiplier transforms in the
adjoint of the isometry group. A second Lagrange multiplier is added
to give the correct boundary conditions on the gauge parameters, and as 
we have seen in the beginning it finally drops out, (here we start with
the Neumann string). Then rewriting the action such that there are no 
derivatives of the isometry gauge fields and factorizing the $\Omega $
integral into a bulk part and a boundary part will give the dual bulk action
and Dirichlet conditions on the dual coordinates $\Lambda_I$
\begin{equation}
\Lambda_I + C_I = 0 \,\,\, \mbox{on} \, \partial \Sigma ,
\end{equation}
being covariant with respect to the isometry group. The gauge fixing is done
by fixing the $X^m$, and in general one has to fix some of the $\Lambda_I$ as
well in order to remove a residual gauge symmetry.
\section{Conclusions}
We have seen that T-duality in open string models interchanges
Neumann with Dirichlet boundary conditions. Specifically, Neumann
boundary conditions correspond to free varying ends of the
string and are not imposed as an external constraint whereas
Dirichlet conditions arise as constraints imposed by a delta
function. We argued that a canonical transformation gives the
same result as long as the generating functional has no
additional boundary contributions. Since this is in
agreement with the Ro\v{c}ek-Verlinde \cite{ro-ve} method it
seems to be natural not to add boundary contributions to
the generating functional.
 
We finish this talk by mentioning some open problems. One problem is,
as in the closed string case, the treatment of global issues.
Another problem is the dilaton shift. In a general sigma model
with boundary there could be two dilatons, one coupling to $R^{(2)}$ 
in the bulk and another one coupling to the geodesic curvature on the 
boundary. String theory tells us that there is only one dilaton coupling to
the Gauss-Bonnet density. So, a natural guess would be that the dual
dilaton couples also to the Gauss-Bonnet density and receives the same
shift as in the closed string case \cite{quev}. A proof for this guess is
missing, so far. It might be as well possible that the dilaton 
receives (infinite) contributions at the position of the
D-brane \cite{jo-witten} which would result in different
boundary and bulk dilatons.            
\section*{Acknowledgments}
S. F. would like to thank Javier Borlaf, Harald Dorn and
Konstadinos Sfetsos for fruitful discussions. The work of
S. S. is supported by DFG - Deutsche Forschungs Gemeinschaft.
A. K. is supported by the Alexander von Humboldt Foundation and
S. F. is supported by GIF- German Israeli Foundation for
Scientific Research. 

\begin{appendix}
\section{Canonical transformation for the open string - preliminaries}
In the Hamiltonian picture the action for the open string is given by
\begin{equation}
S\left[ X, P\right] = \int_{\Sigma} d^2\sigma \left( {\cal H} 
+ P\dot{X}\right)
+ \int_{\partial\Sigma}d\tau\, \left( h\left( X, P\right) + P \dot{X}\right).
\end{equation}
The equation of motion arising from $\frac{\delta S}{\delta X}$ reads
\begin{equation}
\dot{P} = \frac{\partial {\cal H}}{\partial X} - 
\left(\frac{ \partial {\cal H}}{\partial X^\prime}
\right)^\prime
\end{equation}
in the bulk and
\begin{equation}
\delta X \dot{P}_{|\partial \Sigma} = \left( \frac{\partial h}{\partial X} +
\frac{\partial {\cal H}}{\partial X^\prime}\right) \delta X _{|\partial \Sigma}
\end{equation}
on the boundary.
Consider now an equivalent action $$\tilde{S}\left[ \tilde{X},
\tilde{P}\right].$$
This action is equivalent to the first one provided that
\begin{equation} \label{def}
\delta\left( S - \tilde{S}\right) = \int d\tau\, \frac{d}{d\tau}\delta F\left[X,
 \tilde{X}\right] 
\end{equation} 
defining the generating functional $F$.
Taking into account the equations of motion, (the ones from 
$\delta S/ \delta P$ are not needed explicitly), (\ref{def}) leads to
\begin{equation}
P\delta X - \tilde{P}\delta{\tilde{X}} = \delta F
\end{equation}
also in the case of open strings.
\end{appendix}

\end{document}